\documentclass[10pt,epsf]{article}

\usepackage{graphicx}
\usepackage{indentfirst}
\usepackage{cite}
\usepackage{color}

\begin{document}

\title{\bf{Cosmic Censorship of Rotating Anti-de Sitter Black Hole}}

\date{}
\maketitle

\begin{center}
\author{Bogeun Gwak}$^a$\footnote{rasenis@sogang.ac.kr}, \author{Bum-Hoon Lee}$^{a, b, c, d}$\footnote{bhl@sogang.ac.kr}\\
\vskip 0.25in
$^{a}$\it{Center for Quantum Spacetime, Sogang University, Seoul 04107, Korea}\\
$^{b}$\it{Department of Physics, Sogang University, Seoul 04107, Korea}\\
$^{c}$\it{Asia Pacific Center for Theoretical Physics, Pohang 37673, Korea}\\
$^{d}$\it{Department of Physics, Postech, Pohang 37673, Korea}
\end{center}

\vskip 0.6in

\abstract
{We test the validity of cosmic censorship in the rotating anti-de Sitter black hole. For this purpose, we investigate whether the extremal black hole can be overspun by the particle absorption. The particle absorption will change the mass and angular momentum of the black hole, which is analyzed using the Hamilton-Jacobi equations consistent with the laws of thermodynamics. We have found that the mass of the extremal black hole increases more than the angular momentum. Therefore, the outer horizon of the black hole still exists, and cosmic censorship is valid.}

\thispagestyle{empty}
\newpage
\setcounter{page}{1}
\section{Introduction}
Black holes have been investigated in the theories of the gravity and cosmology. One of interesting properties of black holes is the thermodynamics. The black hole has irreducible mass, which increases in a natural process\,\cite{Christodoulou:1970wf,Christodoulou:1972kt}. It will be related to the entropy of the black hole. The irreducible mass is distributed to the surface of the black hole horizon\,\cite{Smarr:1972kt}. The black hole also has a reducible energy, such as rotational energy for rotating black holes or the electric energy of charged black holes. This energy can decrease under specific process such as Penrose process\,\cite{Bardeen:1970zz,Penrose:1971uk}. As a thermodynamic system, the temperature of a black hole can be defined by Hawking radiation \,\cite{Hawking:1974sw,Hawking:1976de}, called the Hawking temperature. The entropy of a black hole is proportional to the area of the black hole horizon, known as Bekenstein-Hawking entropy\,\cite{Bekenstein:1973ur,Bekenstein:1974ax}. Under the definitions of temperature and entropy, we can construct the laws of thermodynamics.

With the Higgs particle discovered at the Large Hadron Collider (LHC)\,\cite{ATLAS:2012ae,Chatrchyan:2012tx}, black holes may play an important role in the early universe. The running of the coupling in the Higgs potential suggests that the present universe may be metastable. The lifetime decaying into true vacua is large enough compared with the age of the universe due to the large energy barrier\,\cite{Coleman:1977py,Callan:1977pt,Coleman:1980aw}. However, the inhomogeneities can effectively lower the energy barrier and reduce the lifetime up to millions of Planck times\,\cite{Burda:2015isa}. These inhomogeneities can be generated from gravitational impurities such as black holes. Anti-de Sitter(AdS) black holes are relevant considering the negative energy of true vacuum\,\cite{Burda:2015yfa}.

AdS black holes are also important in the anti-de Sitter/conformal field theory(AdS/CFT) correspondence\,\cite{Witten:1998zw,Maldacena:1997re,Gubser:1998bc,Witten:1998qj,Aharony:1999ti}. According to the AdS/CFT correspondence, the theory of gravity as defined in bulk AdS spacetime is related to a CFT on the AdS boundary. The dual CFT can be interpreted with the finite temperature theory, in which the temperature corresponds to that of the AdS black hole. The black hole entropy is interpreted through the AdS/CFT correspondence in terms of the microstates\,\cite{York:1983zb,Zurek:1985gd,Frolov:1993ym,Larsen:1995ss}. The microscopic origin of the black hole entropy is accurately obtained\,\cite{Strominger:1997eq} in the (2+1)-dimensional Ba\~{n}ados-Teitelboim-Zanelli (BTZ) black hole\,\cite{Banados:1992wn,Banados:1992gq} by the Cardy formula\,\cite{Cardy:1986ie,Brown:1986nw,Coussaert:1993jp}. The stability of the AdS black hole is related to the dual CFT theory. For example, The RN-AdS black hole is unstable under the presence of charged scalar fields in parameter range\,\cite{Gubser:2008px}; this corresponds to the superconducting instability\,\cite{Hartnoll:2008vx,Hartnoll:2008kx}. In addition, much of this correspondence is related to the condensed matter theory (CMT), called the AdS/CMT correspondence. The charged AdS black holes\,\cite{Martinez:1999qi} have the dual CFTs on their AdS boundaries\,\cite{Maity:2009zz,Chaturvedi:2013ova}, interpreted as holographic superconductors\,\cite{Ren:2010ha,Jensen:2010em,Andrade:2011sx,Haehl:2013hoa,Chang:2014jna}.

The AdS black hole has a singularity inside of the horizon. A naked singularity is not allowed according to cosmic censorship\,\cite{Penrose:1969pc}. For rotating black hole, overspinning beyond the extremal bound, it becomes a naked singularity. Cosmic censorship is invalid in this case. As the source of gravitational impurities, cosmic censorship should be valid in rotating AdS black holes. In 4-dimensional black holes, cosmic censorship is known to be valid in the Kerr black hole under particle absorption\,\cite{Wald:1974ge}. The near-extremal Kerr black hole can be overspun by a particle\,\cite{Jacobson:2009kt}. With self-force, cosmic censorship is valid in the Kerr black hole\,\cite{Barausse:2010ka,Barausse:2011vx,Colleoni:2015afa,Colleoni:2015ena}. In addition, a charged particle can overcharge the near-extremal Reissner-Nordstr\"{o}m (RN) black hole beyond the extremality\,\cite{Hubeny:1998ga}. However, cosmic censorship is still valid in the RN black hole with consideration of back-reaction\,\cite{Isoyama:2011ea}. The validity of cosmic censorship is tested in the several rotating black holes\,\cite{Myers:1986un,BouhmadiLopez:2010vc,Doukas:2010be,Saa:2011wq,Gwak:2011rp,Gao:2012ca} and the anti-de Sitter black holes (AdS)\,\cite{Rocha:2014gza,Rocha:2014jma,McInnes:2015vga} including (2+1)-dimensional black holes\cite{Rocha:2011wp,Gwak:2012hq}.

In this paper, the validity of cosmic censorship is tested in the 4-dimensional rotating AdS black hole. We will investigate whether the extremal black hole can be overspun through particle absorption. We will show that the particle equations of motions should satisfy the laws of thermodynamics for the validity of cosmic censorship under the process. This guarantees the mass increase of the extremal black hole larger than the angular momentum. Thus, cosmic censorship is valid.

The paper is organized as follows. In section~\ref{sec2}, we introduce the rotating AdS black hole solution. In section~\ref{sec3}, we redefine the particle energy in the particle energy equation and prove the laws of thermodynamics in the particle absorption. In section~\ref{sec4}, we will prove the validity of cosmic censorship for the extremal black hole. In section~\ref{sec5}, we briefly summarize our results.

\section{The Rotating AdS Black Hole}\label{sec2}
The rotating AdS black hole is the solution to 4-dimensional Einstein gravity with the negative cosmological constant. The black hole has mass parameter $M$ and spin parameter $a$. The metric is
\begin{eqnarray}\label{eq:metric}
&&ds^2=-\frac{\Delta_r}{\rho^2}\left(dt-\frac{a\sin^2\theta}{\Xi} d\phi\right)^2+\frac{\rho^2}{\Delta_r}dr^2+\frac{\rho^2}{\Delta_\theta}d\theta^2+\frac{\Delta_\theta\sin^2\theta}{\rho^2}\left(a\,dt-\frac{r^2+a^2}{\Xi}d\phi\right)^2\,,\\
&&\rho^2=r^2+a^2\cos^2\theta\,,\quad\Delta_r=(r^2+a^2)(1+\frac{r^2}{\ell^2})-2Mr\,,\quad\Delta_\theta=1-\frac{a^2}{\ell^2}\cos^2\theta\,,\,\,\Xi=1-\frac{a^2}{\ell^2}\,,\nonumber
\end{eqnarray}
where the cosmological constant $\Lambda$ is $-3/\ell^2$ in terms of the AdS radius $\ell$. The function $\Xi$ is defined as a positive value. The outer horizon $r_h$ is located at $\Delta_h\equiv\Delta_r\big|_{r=r_h}=0$. The angular velocity $\Omega_h$ at the black hole outer horizon is $\Omega_h=\frac{a\Xi}{r_h^2+a^2}$. However, the metric still has a non-zero angular velocity $\Omega_\infty=-\frac{a}{\ell^2}$ at the boundary, $r\rightarrow \infty$. Therefore, the angular velocity at the horizon for the non-rotating frame observer is
\begin{eqnarray}
\Omega=\Omega_h-\Omega_\infty=\frac{a\left(1+\frac{r_h^2}{\ell^2}\right)}{r_h^2+a^2}\,.
\end{eqnarray}
The black hole mass $M_B$ and angular momentum $J_B$ are defined in the non-rotating frame\,\cite{Gibbons:2004ai,Caldarelli:1999xj,Aliev:2007qi,Chen:2010jj}
\begin{eqnarray}\label{mass07}
M_B=\frac{M}{\Xi^2}\,,\quad J_B=\frac{aM}{\Xi^2}\,,
\end{eqnarray}
where $G=1$ in this paper. The black hole horizon area $\mathcal{A}_h$ and Bekenstein-Hawking entropy $S_{BH}$ are
\begin{eqnarray}\label{eq:ent07}
S_{BH}=\frac{1}{4}\mathcal{A}_h=\frac{\pi\left(r_h^2+a^2\right)}{\Xi}\,,
\end{eqnarray}
and the Hawking temperature is
\begin{eqnarray}
T_H=\frac{r_h\left(1+\frac{a^2}{\ell^2}+\frac{3r_h^2}{\ell^2}-\frac{a^2}{r_h^2}\right)}{4\pi\left(r_h^2+a^2\right)}\,.
\end{eqnarray}

\section{The Thermodynamics under Particle Absorption}\label{sec3}
We now test the validity of cosmic censorship through a particle that should satisfy the first and second laws of thermodynamics. The black hole changes infinitesimally due to energy and momenta of the particle entering into the black hole horizon. To set the particle momenta as control parameters, the first-order equations of motions are obtained using the Hamilton-Jacobi method\,\cite{Carter} with the separability\,\cite{Vasudevan:2005js}. Hamiltonian with the particle momenta $p_\mu$ is written
\begin{eqnarray}
\mathcal{H}=\frac{1}{2}g^{\mu\nu} p_\mu p_\nu\,.
\end{eqnarray}
The Hamilton-Jacobi action with the particle mass $m$ is constructed as
\begin{eqnarray}\label{eq:nsingle1}
S=\frac{1}{2}m^2\lambda-Et+L\phi+S_r(r)+S_\theta(\theta)\,,
\end{eqnarray}
where the conserved quantities are $E$ and $L$, corresponding to the particle energy and angular momentum. We can find all of geodesics from Eq.~(\ref{eq:nsingle1}). The geodesic equations for the radial momentum $p^r$ and $\theta$-directional angular momentum $p^\theta$ are given as
\begin{eqnarray}
&&p^r\equiv\frac{d r}{d\lambda}=\frac{\Delta_r}{\rho^2}\sqrt{R(r)}\,,\quad\,p^\theta\equiv\frac{d r}{d\lambda}=\frac{\Delta_\theta}{\rho^2}\sqrt{\Theta(\theta)}\,,\\
&&R(r)=\frac{\mathcal{K}}{\Delta_r}+\frac{1}{\Delta_r^2}\left(a \Xi L-\left(r^2+a^2\right)E\right)^2-\frac{m^2r^2}{\Delta_r}\,,\nonumber\\
&&\Theta(\theta)=-\frac{\mathcal{K}}{\Delta_\theta}-\frac{1}{\Delta_\theta^2 }\left(\Xi L\csc\theta-aE\sin\theta\right)^2-\frac{m^2a^2\cos^2\theta}{\Delta_\theta}\,.\nonumber
\end{eqnarray}
These equations are united to construct the particle energy equation by eliminating the separate constant $\mathcal{K}$. The particle energy is obtained from
\begin{eqnarray}
\label{energyeq2}
\alpha E^2 +2\beta E + \gamma=0\,,
\end{eqnarray}
where the coefficients are
\begin{eqnarray}
&&\alpha=\frac{(r^2+a^2)^2}{\Delta_r}-\frac{a^2\sin^2\theta}{\Delta_\theta}\,,\quad\beta=-\frac{(r^2+a^2)(aL\Xi)}{\Delta_r}+\frac{aL\Xi}{\Delta_\theta}\,,\\
&&\gamma=-\frac{\left(p^r\right)^2\rho^4-a^2 L^2 \Xi^2}{\Delta_r}-\frac{\left(p^\theta\right)^2\rho^4+L^2\Xi^2\csc^2\theta}{\Delta_\theta}-m^2\rho^2\,.\nonumber
\end{eqnarray}
The positive-root solution is taken in Eq.~(\ref{energyeq2}) to describe the future-forwarding particle\,\cite{Christodoulou:1970wf,Christodoulou:1972kt}\,. Now, we write the particle energy in terms of the particle momenta at a given location. The conserved quantities of the particle will be absorbed to those of the black hole when the particle passes through the black hole horizon. Solved at the horizon, the particle energy $E_h$ is
\begin{eqnarray}
\label{eq:NminE}
E_h=\frac{a\Xi}{r_h^2+a^2}L+\frac{\rho_h^2}{r_h^2+a^2}|p^r|\,,
\end{eqnarray}
where $\rho^2\big{|}_{r=r_h}=\rho^2_h$. The control parameters are only $p^r$ and $L$ in Eq.~(\ref{eq:NminE}). For the asymptotically flat limit such as $\ell\rightarrow \infty$, this becomes the equation of the Kerr black hole case\cite{Gwak:2011rp}. However, it does not satisfy the second law of thermodynamics in this AdS case. To resolve this problem, the energy is redefined by adding the effect of the boundary. At the limit of $r\rightarrow \infty$, the energy a particle moving on the $\theta=\pi /2$ plane is obtained as
\begin{eqnarray}
E_\infty=-\frac{a}{\ell^2}L+\sqrt{\frac{L^2}{\ell^2}+\frac{(p^r)^2}{\Xi}+\frac{r^2 m^2}{\ell^2\Xi}}\,.
\end{eqnarray}
The spin parameter of the black hole still contributes to the energy of the massless particle with $p^r=0$. The energy contribution is   
\begin{eqnarray}
\hat{E}=-\frac{a}{\ell^2}L\,.
\end{eqnarray}
Note that the ratio of the mass and angular momentum is the same as the angular velocity of the rotating AdS boundary\,$\Omega_\infty$\,\cite{Gibbons:2004ai,Caldarelli:1999xj}. The real particle energy $E_{P}$ is set to no spin contribution of the black hole at infinity, such as in a non-rotating frame, so
\begin{eqnarray}
\label{energycorrect}
E_{P}=E_h-\hat{E} = P_J L+P_P|p^r|\,,\quad P_J=\frac{a\left(1+\frac{r_h^2}{\ell^2}\right)}{r_h^2+a^2}=\Omega\,,\quad P_P=\frac{\rho_h^2}{r_h^2+a^2}\,,
\end{eqnarray}
where the coefficient $P_J$ is the same as the angular velocity $\Omega$ of the black hole in a non-rotating frame. Under the particle energy in Eq.~(\ref{energycorrect}), the corresponding black hole mass and angular momentum infinitesimally change by the particle
\begin{eqnarray}
E_{P}=\delta M_B\,,\,\,\,\,\,L=\delta J_B\,.
\end{eqnarray}
The change in the black hole mass is explicitly described as
\begin{eqnarray}
\label{eq:delta_M}
\delta M_B=P_J\delta J_B+P_P|p^r|\,,
\end{eqnarray}
which will guarantee the second law of thermodynamics. The change of the black hole mass may consist of the reducible energy from angular momentum and the irreducible energy from radial momentum in Eq.~(\ref{eq:delta_M}). The irreducible mass $M_{ir}$ is integrated out from removing the rotational energy from the change in black hole mass
\begin{eqnarray}
M_{ir}=\sqrt{\frac{r_h^2+a^2}{\Xi}}\,,\,\,\,\,\,\delta M_{ir}=\frac{2\rho^2_h}{\dot{D}M_{ir}}|p^r|\geq 0\,,\,\,\,\,\,\dot{D}=\frac{\partial \Delta_h}{\partial r_h}=-2M+\frac{2r_h\left(r_h^2+a^2\right)}{\ell^2}+2r_h\left(1+\frac{r_h^2}{\ell^2}\right)\,.
\end{eqnarray}
The square of the irreducible mass is proportional to the black hole entropy $S_{BH}$ in Eq.~(\ref{eq:ent07}). The irreducible mass is only dependent on the particle radial momentum. However, the rotational energy depends on the angular momentum sign $L$, so it stands as a reducible mass. Therefore, the irreducible mass always increases when a particle falls into the black hole. This  guarantees the second law of thermodynamics in particle absorption. To prove this, we perturb the entropy in Eq.~(\ref{eq:ent07}) by the particle energy and momenta constrained in Eq.~(\ref{eq:delta_M}). The change in the entropy is obtained
\begin{eqnarray}
&&\delta S_{BH}=S_{\delta J}\,L+S_{\delta P}|p^r|\,,\,\,\,\,\,S_{\delta J}=P_JS_M-\frac{D_MP_J\dot{S}}{\dot{D}}+S_J-\frac{D_J\dot{S}}{\dot{D}}\,,\,\,\,\,\,S_{\delta P}=P_PS_M-\frac{D_MP_P\dot{S}}{\dot{D}}\,,\nonumber\\
&&S_M=\frac{\partial S_{BH}}{\partial M_B}=-\frac{2a^2\pi\left(r_h^2+a^2\right)}{M\ell^2}-\frac{2a^2\pi\Xi}{M}\,,\,\,\,\,\,S_J=\frac{\partial S_{BH}}{\partial J_B}=\frac{2a\pi\left(r_h^2+a^2\right)}{M\ell^2}+\frac{2a\pi\Xi}{M}\,,\\
&&\dot{S}=\frac{\partial S_{BH}}{\partial r_h}=\frac{2\pi r_h}{\Xi}\,,\,\,\,\,\,D_M=\frac{\partial \Delta_h}{\partial M_B}=-\frac{8a^2 r_h \Xi}{\ell^2} -2r_h \Xi^2 -\frac{2a^2\left(1+\frac{r_h^2}{\ell^2}\right)\Xi^2}{M}\,,\nonumber\\
&&D_J=\frac{\partial \Delta_h}{\partial J_B}=\frac{8a r_h \Xi}{\ell^2} +\frac{2a\left(1+\frac{r_h^2}{\ell^2}\right)\Xi^2}{M}\,.\nonumber
\end{eqnarray} 
This is reduced to the simple equation
\begin{eqnarray}
\label{eq:delent}
\delta S_{BH}=\frac{4\pi \rho_h^2}{\dot{D}}|p^r|\geq 0\,.
\end{eqnarray}
The function $\dot{D}$ is zero at the extremal black hole and positive at the outer horizon of the non-extremal one, so that the total sign of the equation is positive. Therefore, the black hole entropy always increases when a particle is absorbed into the black hole. It proves the second law of thermodynamics under the process. Rewritten from Eq.~(\ref{eq:delta_M}) using Eq.~(\ref{eq:delent}), the change in the black hole mass is
\begin{eqnarray}
\delta M_B = \Omega \,\delta J_B + T_H \,\delta S_{BH}\,, 
\end{eqnarray}
which is the first law of thermodynamics. Therefore, the particle satisfies the laws of thermodynamics.

\section{The Validity of Cosmic Censorship}\label{sec4}
Now, we will prove the extremal black hole for the validity of cosmic censorship under the absorption of this particle. The extremal black hole horizon $r_h$ is located at the minimum point of the function $\Delta_r(r)$\,. Before the particle absorption, the initial extremal black hole horizon satisfies
\begin{eqnarray}
\Delta_h=0\,,\,\,\,\,\,\dot{D}=0\,.
\end{eqnarray}
The black hole mass and angular momentum infinitesimally change after the particle absorption, so the function $\Delta_r(r)$ has a different minimum value. If the positive minimum value of the function $\Delta_r(r)$ is positive, then there does not exist the horizon, and cosmic censorship is invalid. If the function $\Delta_r(r)$ has a negative or zero minimum value, the horizon still exists, which proves the validity of cosmic censorship. The particle changes the minimum point location to $r_h+\delta r_e$,
\begin{eqnarray}
\delta r_e=-\frac{\dot{D}_M P_J+\dot{D}_J}{\ddot{D}}L-\frac{\dot{D}_M P_P}{\ddot{D}}|p^r|\,,
\end{eqnarray}
where
\begin{eqnarray}
&&\ddot{D}=\frac{\partial \dot{D}}{\partial r_h}=2\left(1+\frac{r_h^2}{\ell^2}\right)+\frac{8 r_h^2}{\ell^2}+\frac{2\left(r_h^2+a^2\right)}{\ell^2}\,,\quad \dot{D}_M=\frac{\partial \dot{D}}{\partial M_B}=-\frac{8a^2\Xi}{\ell^2}-2\Xi^2-\frac{4a^2 r_h \Xi^2}{M\ell^2}\,,\nonumber\\
&&\dot{D}_J=\frac{\partial \dot{D}}{\partial J_B}=\frac{8a\Xi}{\ell^2}+\frac{4a r_h \Xi^2}{M\ell^2}\,.\nonumber
\end{eqnarray}
The minimum point $r_h+\delta r_e$ may no longer be the horizon. The particle absorption changes the minimum value of the function $\Delta_r(r)$ as
\begin{eqnarray}
\Delta_r(r_h+\delta r_e)=D_M\delta M_B+D_J \delta J_B+ \dot{D}\delta r_e=D_M P_P \,|p^r|\,.
\end{eqnarray}
The minimum value is always negative for the extremal black hole, because the coefficient $D_M$ is negative for the black hole. Therefore, this guarantees the validity of cosmic censorship. Also, the minimum is only one extremum of the function $\Delta_r(r)$, because the second derivative of the function $\Delta_r(r)$ with respect to $r$ is always positive. The function $\Delta_r(r)$ gives a positive value at $r=0$, so the negative minimum value provides two horizons. One is outer horizon. The other is inner horizon. This implies that the outer horizon of the black hole still exists, and the extremality is broken for non-zero radial momentum. Note that this conclusion does not depend on the angular momentum of the particle. The non-extremality after the particle absorption implies that the black hole mass increases more than the angular momentum of the extremal black hole. In conclusion, the extremal rotating AdS black hole becomes a non-extremal one, and cosmic censorship is valid under the particle absorption.

\section{Summary}\label{sec5}
Cosmic censorship is valid in the rotating AdS black hole under particle absorption. Using particle energy and momenta, we have investigated whether the extremal black hole can be overspun or not. We have redefined the particle energy to remove the contribution of the black hole rotation at the spacetime boundary. Under the redefinition, the particle satisfies not only the equations of the motions, but also the laws of thermodynamics. After the particle absorption, the extremal black hole mass increases more than the angular momentum of the black hole. In other words, the final state has one more horizon than the initial one, hence it becomes non-extremal, and cosmic censorship is valid.\\

{\bf Acknowledgments}

We would like to thank Gungwon Kang for helpful discussions and comments. This research was supported by Basic Science Research Program through the National Research Foundation of Korea(NRF) funded by the Ministry of  Science, ICT \& Future Planning(NRF-2015R1C1A1A02037523) and the National Research Foundation of Korea(NRF) grant funded by the Korea government(MSIP)(No.~2014R1A2A1A010).

\end{document}